\documentclass[11pt]{article}
\usepackage{amsmath}
\usepackage{amsfonts}
\usepackage{amssymb}
\usepackage{mathrsfs}
\usepackage{dblfloatfix}
\usepackage{float}
\usepackage[margin=2.5cm]{geometry}
\begin{document}

\title{Vacuum stability and the Cholesky decomposition}
\author{James Unwin\thanks{Electronic address: unwin@maths.ox.ac.uk}
\\
\\
 {\normalsize \it  Mathematical Institute, University of Oxford,}\\
 {\normalsize \it  24-29 St. Giles', Oxford, OX1 3LB, United Kingdom}}
\date{}

\maketitle

\abstract{
We discuss how the Cholesky decomposition may be used to ascertain whether a critical point of the field theory scalar potential provides a stable vacuum configuration. We then use this method to derive the stability conditions in a specific example.}

\section{Introduction}
\label{intro}
Understanding the structure of the vacuum is of central importance in quantum field theory. Local minima of the  effective quantum potential correspond to stable vacuum configurations of the system and physical states in the quantum theory are identified with fluctuations around a given minimum.

In this note we shall introduce the Cholesky decomposition and present a method for using the Cholesky decomposition of the Hessian of the potential to ascertain the stability of a given vacuum configuration.
Whilst approaches of this manner are familiar in some mathematical disciplines, it does not seem to be widely known in theoretical physics.

This technique should be of value in the study of vacuum stability in models of interest in particle physics.
In particular, the study of the vacuum structure of gauge mediated supersymmetry breaking models, where stability analysis is often complicated due to the number of scalar fields entering in the potential \cite{Giudice}.

Firstly, we shall present a summary of the relevant notions from linear algebra and discuss their use in demonstrating vacuum stability. To illustrate this approach, in the latter half of the paper, we shall apply these techniques to a particular example for which vacuum stability has not yet been shown. Additional areas for potential application of this approach are suggested in the concluding remarks.

\section{The Cholesky decomposition}
\label{sec:1}

Given a general scalar potential \(V=V(\phi_1,\cdots,\phi_n)\), a stable (degenerate) vacuum configuration of the system is a (semi)stable critical point of \(V\). That is a set of parameters for which
\begin{equation}
\frac{\partial V}{\partial \phi_i}=0 \quad \mathrm{for} \quad i=1,\cdots, n.
\label{e1}
\end{equation}
The Hessian \(H\) of a potential \(V(\phi_1,\cdots,\phi_n)\) is defined to be the matrix of second derivatives of the potential with respect to each of the fields it depends upon:
\begin{equation}
H=\left(\begin{array}{cccc}
\frac{\partial V}{\partial \phi_1 \partial \phi_1 } &
\cdots &
\frac{\partial V}{\partial \phi_1 \partial \phi_n } \\
\vdots &
\ddots &
\vdots\\
\frac{\partial V}{\partial \phi_n \partial \phi_1 } &
\cdots &
\frac{\partial V}{\partial \phi_n \partial \phi_n }
\end{array}\right).
\end{equation}

The Hessian is a useful object since the stability of a given critical point can be inferred from the properties of the matrix. Suppose that the point \(p\) is a critical point of \(V\), then it is (semi)stable if and only if the Hessian of \(V\)  at \(p\) is positive-(semi)definite \cite{hessian}.
If the Hessian \(H\) is singular at \(p\), then using row operations \(H\) may be separated into a nonsingular symmetric submatrix \(M\) and a set of zero vectors. If the submatrix \(M\) is positive-definite, then \(H\) is positive-semidefinite.

For a Hermitian or real symmetric matrix \(A\) the following three statements are equivalent \cite{cholesky}:
\renewcommand{\theenumi}{(\roman{enumi})}
\begin{enumerate}
\item \(A\) is positive-definite, i.e.
\begin{equation}
\langle \boldsymbol{x}, A\boldsymbol{x} \rangle > 0
\end{equation}
 for all nonzero vectors \(\boldsymbol{x}\).
\item All of the eigenvalues of \(A\) are positive.
%\item There exists a nonsingular matrix \(A\) such that
%\begin{equation}
%M=A^\dagger A.
%\end{equation}
\item \(A\) can be uniquely decomposed as follows
\begin{equation}
A=C^\dagger C
\end{equation}
where \(C\) is an upper triangular matrix with real strictly positive diagonal entries.
\label{iii}
\end{enumerate}

The factorisation of \(A\) into \(C\) and its conjugate in \ref{iii} is  referred to as the Cholesky decomposition of \(A\) and may be considered as the matrix analogue of the square root operation for scalar numbers.

Additionally, the three statements below are also equivalent to each other \cite{cholesky}:
\renewcommand{\theenumi}{(\alph{enumi})}
\begin{enumerate}
\item \(A\) is positive-semidefinite, i.e.
\begin{equation}
\langle \boldsymbol{x}, A\boldsymbol{x} \rangle \geq 0
\end{equation}
 for all nonzero vectors \(\boldsymbol{x}\).
\item All of the eigenvalues of \(A\) are non-negative.
\item There exists a (possibly singular) matrix \(B\) such that
\begin{equation}
A=B^\dagger B.
\end{equation}
\end{enumerate}

Whilst one can ascertain whether a matrix is positive-(semi)definite by examining its eigenvalues, finding analytically the eigenvalues of \(H\) evaluated at \(p\) becomes increasingly more difficult as the dimension of the parameter space grows.
Demonstrating the existence of the Cholesky decomposition provides an alternative method to finding the eigenvalues of the Hessian and can often be more practical to compute.

Since the Hessian is symmetric, if it is nonsingular at a given critical point \(p\) then it can be factorised as \(H=U^\dagger U\), where \(U\) is an upper triangular matrix with symbolic entries. The requirement that the diagonal entries of \(U\) be real and positive provides the stability conditions for the vacuum configuration. Note that when the diagonal entries are real and positive, the matrix \(U\) is the Cholesky decomposition of \(H\).
%In which case, since \(H\) admits a Cholesky decomposition, the critical point is stable.
If the Hessian is singular at \(p\), then one must inspect the nonsingular submatrix \(M\).
Similarly, one may factorise \(M=U^\dagger U\) and the vacuum is semistable if the diagonal entries of \(U\) are real and positive.

Note that the dimension of the singular subspace gives the number of massless states, some of which may be Nambu-Goldstone bosons, but others correspond to moduli (`flat-direction' fields in the supersymmetric case).

For a given Hermitian or real symmetric matrix \(A\) with symbolic entries, there are algorithms \cite{alg}, based on iterative elementary row operations, through which one may obtain the factorisation \(A=U^\dagger U\). In the case where \(A\) is positive-definite this provides the Cholesky decomposition. Moreover, most standard mathematical computing packages can perform this operation via preprogrammed functions \cite{math}.

\section{A particular example}
\label{sec:2}

We shall illustrate here the method outlined in Section \ref{sec:1} by deriving the stability conditions for the vacuum configuration of a model proposed by Chan and Tsou \cite{chan}, related to the Standard Model. The problem requires the analysis of a 10\(\times\)10 Hessian and thus provides a sufficiently complicated example to demonstrate the utility of our method.

The scalar content of the theory under consideration is as follows:
one complex scalar field \(\varphi\) transforming in the fundamental representation of su(2) and three complex scalar fields \(\phi^i\) (\(i=1,2,3\)) transforming in the fundamental representation of su(3).
The scalar potential of the model has the following form
\begin{equation}
\begin{aligned}
 V &= -\mu_w |\varphi|^2 +\lambda_w|\varphi|^4     -\mu_s\sum_{a=1}^3\sum_{i=1}^3|\phi_a^i|^2\\
&\hspace{2mm}
+\lambda_s\left(\sum_{a=1}^3\sum_{i=1}^3|\phi_a^i|^2\right)^2
+\kappa_s\sum_{a,b=1}^3\left|\phi_a^i\cdot\phi_b^i{}^*\right|^2 \\
&\hspace{2mm} +\nu_1|\varphi|^2\sum_{a=1}^3\sum_{i=1}^3|\phi_a^i|^2-\nu_2|\varphi|^2\sum_{a=1}^3\left|{\boldsymbol{\alpha}}_i\cdot\phi_a^i{}^*\right|^2.
\label{35}
\end{aligned}
\end{equation}
The vector \(\boldsymbol{\alpha}\) is a global vector and is contracted against the multiplicity index \(i\) of the scalar fields \(\phi^i\). A global symmetry of the model allows one to fix the orientation of \(\boldsymbol{\alpha}\) and henceforth we shall work with \(\boldsymbol{\alpha}=(1,\,0,\,0)^T\).
Following \cite{chan}, we fix the symmetries of the theory and subsequently parametrise the scalar fields thus
\begin{equation}
\begin{aligned}
%\label{v1}
\varphi_r &=\left(
\begin{array}{cc}
\zeta_w\\
0
\end{array}
\right),\\
{\phi_a^1} &=\left(
\begin{array}{ccc}
X \cos \delta_1  \\
0\\
0\\
\end{array}
\right),\\
{\phi_a^2} &=\left(
\begin{array}{ccc}
 X \sin\delta_1\sin\gamma \, e^{i\chi_3} \\
 Y \cos\delta_2 \\
0
\end{array}
\right),\\
{\phi_a^3} &=\left(
\begin{array}{ccc}
 X \sin\delta_1\cos\gamma \, e^{i\chi_2}\\
 Y\sin\delta_2 \, e^{i\chi_1}\\
 Z\\
\end{array}
\right),
\label{v2}
\end{aligned}
\end{equation}
where the above variables are real and obey the following relationship
\begin{equation}
X^2+Y^2+Z^2=\zeta_s^2.
\label{sum}
\end{equation}
The quantities \(\zeta_w\) and \(\zeta_s\) are identified with the absolute lengths of the sets of scalar fields:
\begin{equation}
\begin{aligned}
\zeta_w &=\sqrt{|\varphi|^2}, \\
 \zeta_s &=\sqrt{|\phi^1|^2+|\phi^2|^2+|\phi^3|^2}.
\end{aligned}
\end{equation}
Expanding the potential \(V\) in terms of equations (\ref{v2}) we obtain
\begin{equation} \begin{aligned}
V=&\label{Pot}
-\mu_w\zeta_w^2 + \lambda _w \zeta_w^4\\
&-\mu _s\Big(X^2+Y^2+Z^2\Big) +\lambda _s
   \Big(X^2+Y^2+Z^2\Big)^2\\
&+  \kappa _s \Big[
X^4+Y^4+Z^4\\
&\hspace{1mm} + 2X^2Y^2 \sin ^2 \delta _1 \cos ^2 \delta _2 \sin ^2 \gamma \\
&\hspace{1mm} + 2X^2Y^2 \sin ^2 \delta _1 \sin ^2 \delta _2 \cos ^2 \gamma\\
&\hspace{1mm} + X^2 Y^2 \sin ^2\delta _1 \sin 2\gamma \sin 2 \delta_2 \cos(\chi_2-\chi_1-\chi_3)\\
&\hspace{1mm} + 2X^2Z^2 \sin ^2 \delta _1 \cos ^2
 \gamma
 + 2Y^2Z^2 \sin ^2 \delta _2\Big] \\
 & +\nu _1 \zeta_w^2\Big(X^2+Y^2+Z^2\Big)
-\nu _2 \zeta_w^2 X^2\cos^ 2 \delta_1\\
\end{aligned}\end{equation}
For convenience the following quantities are defined
\begin{equation}
\begin{aligned}
\Delta_1 &=X^2-Y^2,\\
\Delta_2 &=Y^2-Z^2,\\
\label{del}
R&=\frac{\nu_2\zeta_w^2}{2\kappa_s\zeta_s^2}.
\end{aligned}
\end{equation}
We consider the following critical point of the potential \(V\), which we express in terms of \(\zeta_s\), \(\Delta_1\) and \(\Delta_2\):
\begin{equation}\begin{aligned}
\chi_1 &=\chi_2=\chi_3=  \delta_1 = \delta_2 = \Delta_2  = 0,\\
\Delta_1 &= R,\\
\zeta_w^2 &=
\frac{3\mu _w+(1+2R)\nu_2\zeta _s^2-3\nu_1\zeta _s^2}{6\lambda _w},\\
\zeta _s^2
 &=\frac{3 \mu _s-3\nu_1\zeta _w^2+\nu_2\zeta _w^2}{2\kappa _s+6  \lambda _s}.
 \label{Q2}
\end{aligned}\end{equation}
From equations (\ref{sum}), (\ref{del}) and (\ref{Q2}) we obtain the relationships
\begin{equation}
\begin{aligned}
X &=\zeta_s \sqrt{\frac{1+2R}{3}},\\
Y &=Z=\zeta_s \sqrt{\frac{1-R}{3}}.
\label{XY}
\end{aligned}
\end{equation}
Since the variables \(X,\, Y\) and \(Z\) are real it follows that
\begin{equation}
-\frac{1}{2}\leq R \leq 1.
\end{equation}

The Hessian \(H\) of the system is constructed as a symmetric matrix in the basis:
\[\{X,\, Y,\,  Z,\, \zeta_w,\, \delta_1,\,  \delta_2, \, \gamma, \, \tau_1, \, \tau_2, \, \tau_3\}.\]
The form of the Hessian at the critical point defined by the set of equations (\ref{Q2}) is found to be the following
\begin{equation}
H=8 \cdot \left(\begin{array}{ccc|c}
  N & 0 & 0   & \\
  0 & \frac{X^2}{4}  \left(2 Y^2 \kappa _s+\nu _2 \zeta _w^2\right)   & 0 & \boldsymbol{0}\\
  0 & 0  &  \frac{Y^4 \kappa_s}{2}  & \\
\hline
    & \boldsymbol{0}  &   & \,\boldsymbol{0}_4
\end{array}\right)
\end{equation}
where \(N\) is a 4\(\times\)4 matrix given by
\begin{equation}
N=\left(
\begin{array}{cccc}
  X^2 \left(\kappa _s+\lambda _s\right) &  X Y \lambda _s &  X Y \lambda _s & \frac{ X \zeta _w}{2} \left(\nu _1-\nu _2\right) \\
  X Y \lambda _s &  Y^2 \left(\kappa _s+\lambda _s\right) &  Y^2 \lambda _s &  \frac{ Y \zeta _w \nu _1}{2} \\
  X Y \lambda _s &  Y^2 \lambda _s &  Y^2 \left(\kappa _s+\lambda _s\right) &  \frac{ Y \zeta _w \nu _1}{2} \\
\frac{ X \zeta _w}{2} \left(\nu _1-\nu _2\right) & \frac{ Y \zeta _w \nu _1}{2} & \frac{ Y \zeta _w \nu _1}{2} & \zeta_w^2\lambda_w
\end{array}\right)
\end{equation}
The Hessian is positive-semidefinite if the submatrix \(N\)
is  positive-semidefinite and the other entries of \(H\) are non-negative:
\begin{equation}
\begin{aligned}
 0 & \leq  \frac{X^2}{4}  \left(2 Y^2 \kappa _s+\nu _2 \zeta _w^2\right)  ,\\
 0 & \leq  \frac{Y^4 \kappa_s}{2}.
\end{aligned}
\end{equation}
Using equations (\ref{XY}) it can be seen that these entries are non-negative given the following conditions
\begin{equation}
\begin{aligned}
R &\geq-\frac{1}{2}\\
 \kappa_s &\geq 0.
\label{c1}
\end{aligned}
\end{equation}

Note that for certain values of the variables and parameters, the submatrix \(N\) is singular; for instance:
\[R=1,\,-\frac{1}{2},\quad \zeta_s=0, \quad\mathrm{or}\quad \zeta_w=0.\]
 For brevity we shall consider only the case where  \(N\) is nonsingular, as the special cases where \(N\) is singular can be analysed using the same method and are computationally easier.

Finding the eigenvalues of \(N\) analytically is difficult and has not been previously completed.
Instead we shall utilise the method detailed in Section \ref{sec:1} to analyse the vacuum stability.
The submatrix \(N\) may be factorised as \(N = U^{\dagger}U\),
where the form of \(U\) is given by
%in equation (\ref{d}) at the foot of the page.

\begin{footnotesize}
\begin{equation*}
\hspace{-5mm}
\left(
\begin{array}{cccc}
 \frac{ \zeta _s \sqrt{(1+2R)(\kappa _s+\lambda _s)}}{\sqrt{3}} &  \frac{\zeta _s \lambda_s \sqrt{(1-R) }}{\sqrt{3(\kappa _s+\lambda _s)}} &  \frac{\zeta _s\lambda _s\sqrt{(1-R) }}{\sqrt{3(\kappa _s+\lambda _s)}} & \frac{\zeta _w \left(\nu _1-\nu _2\right)}{2 \sqrt{\kappa _s+\lambda _s}} \\
 0 & \frac{ \zeta _s \sqrt{(1-R)\kappa _s \left(\kappa _s+2 \lambda _s\right)}}{\sqrt{3(\kappa _s+\lambda _s)}} & \frac{\zeta _s\kappa _s \lambda  _s\sqrt{(1-R) }}{\sqrt{3\kappa _s \left(\kappa _s+\lambda _s\right) \left(\kappa _s+2 \lambda _s\right)}}
 & \frac{\zeta _w \left(\kappa _s
   \nu _1+\lambda _s \nu _2\right)}{2 \sqrt{\kappa _s \left(\kappa _s+\lambda _s\right) \left(\kappa _s+2 \lambda _s\right)}} \\
 0 & 0 & \frac{ \zeta _s \sqrt{(1-R)\kappa _s \left(\kappa _s+3 \lambda _s\right)}}{\sqrt{3(\kappa _s+2 \lambda _s)}} & \frac{\zeta _w \left(\kappa _s
   \nu _1+\lambda _s \nu _2\right)}{2 \sqrt{\kappa _s \left(\kappa _s+2 \lambda _s\right) \left(\kappa _s+3 \lambda _s\right)}}
   \\
 0 & 0 & 0 &  \frac{\zeta _w \sqrt{4 \lambda _w \kappa _s^2+\left(12 \lambda _s \lambda _w-3 \nu _1^2+2 \nu _2 \nu _1-\nu _2^2\right) \kappa _s-2 \lambda _s \nu _2^2}}{2 \sqrt{\kappa _s \left(\kappa _s+3 \lambda _s\right)}}
\end{array}
\right)
%\tag{$\star$}
%\label{d}
\end{equation*}
\end{footnotesize}

Recall that the quantities \(\zeta_w\) and \(\zeta_s\) are identified with lengths and thus are defined to be non-negative.
Whence we conclude that the diagonal entries of \(U\) are real and positive given the following further conditions
\begin{equation}\begin{aligned}
\lambda _s & > -\frac{\kappa_s}{3},\\
\lambda _w & > \frac{2 \nu _2^2 \lambda _s +(3 \nu _1^2-2 \nu _1 \nu _2+\nu _2^2)\kappa _s}{4\kappa _s(\kappa _s+3\lambda _s)}.
\label{c2}
\end{aligned}\end{equation}
In which case, by the argument detailed in Section \ref{sec:1}, the Hessian is positive-semidefinite.
Therefore we have shown that the set of semistable vacua is nonempty and furthermore we have derived the provisions, (\ref{c1}) and (\ref{c2}), under which the vacuum configuration is stable.

\section{Concluding remarks}

We have outlined how the Cholesky decomposition may be applied to the Hessian of a system in order to ascertain the stability of a given critical point.
In the example considered above it was not practical to find the eigenvalues of the Hessian analytically and the Cholesky decomposition was key to obtaining the stability conditions for the model under consideration.

Although techniques based upon the Cholesky decomposition are established in some mathematical fields, they have not been widely used in theoretical physics for demonstrating vacuum stability or otherwise.
One interesting application of the Cholesky decomposition was presented in \cite{Leontaris}, in the  analysis of  fermion masses in a wide class of effective low energy models emerging from intersecting D-brane configurations.

Whilst the application of our technique is restricted in this note to the analysis of the Chan-Tsou model, it is evident from our calculation that the approach is quite general. This method is well suited for the study of models of interest in theoretical physics, many of which present a large number of scalar degrees of freedom, and it is expected that the procedure we have presented may find several further applications in particle theory.

Supersymmetric theories are an immediate candidate for applying this technique as they generally involve many complex scalar fields, leading to complicated effective potentials. It is expected that the approach advocated above may be of use in stability analysis and the classification of flat directions in supersymmetric models \cite{Giudice}, \cite{2higgs}, \cite{JMR}.

A further opportunity for application is the stability analysis of various grand unified theories which feature, for phenomenological reasons, scalar fields in high dimensional representations of the gauge group \cite{gut}. Whilst work has been undertaken to identify stable minima in such models \cite{gut2}, there remain open problems which might be approached using the method presented here.\\

\smallskip

\noindent \textit{Acknowledgements.} I would like to thank Tsou Sheung Tsun and Laura Schaposnik for useful discussions. I am grateful also to Fidel Schaposnik and John March-Russell for their comments on a draft of this paper. This work was funded through an EPSRC doctoral training account, with additional support from Pembroke College, Oxford.

\end{document}